\documentclass[12pt]{iopart}
\usepackage{amsmath}
\usepackage{graphicx}
\usepackage{float}
\usepackage{svg}
\usepackage{pdflscape}
\usepackage{lipsum}
\usepackage{booktabs} 
\usepackage{url}

\usepackage{array}
\newcolumntype{L}[1]{>{\raggedright\let\newline\\\arraybackslash\hspace{0pt}}m{#1}}
\newcolumntype{C}[1]{>{\centering\let\newline\\\arraybackslash\hspace{0pt}}m{#1}}
\newcolumntype{R}[1]{>{\raggedleft\let\newline\\\arraybackslash\hspace{0pt}}m{#1}}

\usepackage{pifont}

\begin{document}

\title[Constant Velocity Physical Warp Drive Solution]{Constant Velocity Physical Warp Drive Solution}

\author{Jared Fuchs$^\dagger{}^{1,2}$, Christopher Helmerich${}^{1,2}$, Alexey Bobrick${}^{2,3}$, \ \ \ Luke Sellers${}^{2,4}$, Brandon Melcher${}^{2}$, \& Gianni Martire${}^{2}$}

\address{{${}^{1}$The University of Alabama in Huntsville, 301 Sparkman Drive,
Huntsville, Alabama, 35899, U.S.}}
\address{${}^{2}$Advanced Propulsion Laboratory at Applied Physics, 477 Madison Avenue, New York, 10022, U.S.}
\address{${}^{3}$Technion - Israel Institute of Technology, Physics Department, Haifa 32000, Israel}
\address{${}^{4}$UCLA Department of Physics \& Astronomy,
475 Portola Plaza, Los Angeles, CA 90095, U.S.}

\ead{$^\dagger$ jef0011@uah.edu \& jared@appliedphysics.org}

\vspace{10pt}
\begin{indented}
\item[]August 2023
\end{indented}

\begin{abstract}
Warp drives are exotic solutions of general relativity that offer novel means of transportation. In this study, we present a solution for a constant-velocity subluminal warp drive that satisfies all of the energy conditions. The solution involves combining a stable matter shell with a shift vector distribution that closely matches well-known warp drive solutions such as the Alcubierre metric. We generate the spacetime metric numerically, evaluate the energy conditions, and confirm that the shift vector distribution cannot be reduced to a coordinate transformation. This study demonstrates that classic warp drive spacetimes can be made to satisfy the energy conditions by adding a regular matter shell with a positive ADM mass. 
\end{abstract}

%
%
%
%
%

\section{Introduction}
Warp drive spacetimes, first introduced by Alcubierre \cite{1994CQGra..11L..73A} and later by others
\cite{1999CQGra..16.3973V, 2002CQGra..19.1157N}, offer several unique transportation properties for timelike observers. These properties include the possibility of accelerating through geodesic motion, moving superluminally, or being in regions of modified spacetime, all relative to external inertially moving timelike observers. All of these classic warp drive spacetimes violate some if not all of the energy conditions \cite{1995PhRvD..51.4277F, 2022PhRvD.105f4038S}, and therefore, their construction has been largely considered to be unfeasible. However, recent papers \cite{2022arXiv220100652L, 2021CQGra..38j5009B, 2021CQGra..38o5020F} have suggested that `physical warp drives' that satisfy some or all of the energy conditions could possibly be constructed, reigniting interest in the subject.  
\par 
Due to the complexity of the Einstein equations, progress toward finding physical warp drive solutions through purely analytical means has been slow. In particular, factors that increase the complexity of these equations considerably are non-unit lapse functions and non-flat spatial metrics, both of which have been absent from most previous warp drive solutions and both of which have been argued to be necessary for satisfying the energy conditions \cite{2021CQGra..38j5009B}.  To address this challenge, the computational toolkit, \textit{Warp Factory} \cite{NumericalWarp2023}, was developed to provide numerical methods for exploring warp drive spacetimes more comprehensively. Using this new-found flexibility afforded by \textit{Warp Factory}, we present here a new subluminal constant-velocity warp drive solution with a non-unit lapse and non-flat spatial metric that satisfies all of the energy conditions.

\subsection{Key Features of a Warp Drive Spacetime}
A detailed discussion on the properties of warp drive spacetimes may be found in \cite{NumericalWarp2023}, here we summarize these key features for clarity. Warp drive spacetimes may be viewed as modifications of a background globally hyperbolic and asymptotically-flat spacetime, which contains a generally non-geodesic trajectory $\mathcal{C}_{\rm background}$ connecting two arbitrary points A and B. A warp drive spacetime modifies this background spacetime such that the following is true:

\begin{enumerate}
\item \textit{Geodesic transport: }  
The generally non-geodesic trajectory $\mathcal{C}_{\rm background}$ in the background spacetime maps to a geodesic trajectory $\mathcal{C}_{\rm background}  \rightarrow \mathcal{C}_{\rm warp}  $ in the new warp drive spacetime. In other words, warp drive spacetimes enable passengers to travel between points A and B along a geodesic trajectory $\mathcal{C}_{\rm warp}$. This means the passengers inside the warp drive do not experience local acceleration while being transported\footnote{In the case that a local acceleration is desired, for example, 1g, the statement becomes the `local acceleration of passengers should be limited'.}. For a non-trivial solution, the original trajectory $\mathcal{C}_{\rm background}$ should not be a geodesic, i.e. the passengers should not `already be going' from point A to point B. An example of a non-trivial solution is a passenger in a static background spacetime that is initially at rest at point A (relative to the local frame of rest), is transported to point B, and is then, again, at rest relative to point B. In addition, the warp modification should minimally affect the proper distances between A and B measured along the path, as defined originally in the background spacetime.

\item \textit{Empty passenger region: } The warp drive spacetime has a compact vacuum ($T^{\mu\nu} = 0$)\footnote{Ignoring the mass of the passengers.} passenger region that is free from tidal forces and encloses the passenger trajectory $\mathcal{C}_{\rm warp}$.

\item \textit{A spatially bounded, comoving bubble: } 
The warp drive spacetime has a compact non-vacuum region ($T^{\mu\nu} \neq 0$) that encloses the passenger trajectory $\mathcal{C}_{\rm warp}$ on every spacelike slice. This means the stress-energy distribution required for the geodesic transport does not extend to infinity\footnote{Perhaps, some energy could be radiated to infinity but that energy should be causally connected to the bubble.} and moves along with the transported observers. The requirement of moving with passengers distinguishes warp drive solutions from Krasnikov tubes \cite{1997PhRvD..56.2100E}, for example.

\end{enumerate}

\subsection{Designing Warp Drive Spacetimes}
The transportation element of warp drives is about designing timelike curves for passengers to travel between points A and B in spacetime. In this paper, we will go about developing a warp solution in the following steps:
\begin{enumerate}
    \item Start with a Minkowski background.
    \item Define two points A and B in spacetime.
    \item Define the starting and end conditions for the passengers that will travel between A and B. For example, the passengers might begin at rest at point A and then end at rest at point B. Such starting conditions can be defined w.r.t to an outside observer situated in a Minkowski space.
    \item Define a curve between points A and B that the warp drive and passengers will travel along.
    \item Construct a metric solution that will move passengers within the boundary conditions of (iii) along geodesics matched to the curve in (iv). There are multiple possible metrics that enable these specific geodesics, but only one is needed.
\end{enumerate}
Generally speaking, there are many ways that this can be accomplished. In a 3+1 formalism, the metric is given by:
\begin{equation}
    ds^2 = -\alpha^2 dt^2 + \gamma_{ij} (dx^i + \beta^i dt)(dx^j + \beta^j dt),
\end{equation}
where $\alpha$ is the lapse function, $\beta^i$ is the shift vector and $\gamma_{ij}$ is the spatial metric. We can consider the general geodesic equations in a 3+1 formalism parameterized by the coordinate time \cite{2018ApJS..237....6B} as:
\begin{equation}\label{eq:geodesicEvolution}
\begin{split}
    &\frac{d x^i}{d t}=\gamma^{i j} \frac{u_j}{u^0}-\beta^i \\
    &\frac{d u_i}{d t}=-\alpha u^0 \partial_i \alpha+u_k \partial_i \beta^k-\frac{u_j u_k}{2 u^0} \partial_i \gamma^{j k} \\
    &u^0=\left(\gamma^{j k} u_j u_k+\epsilon\right)^{1 / 2} / \alpha
\end{split}
\end{equation}
where $u^\mu = dx^\mu/d\tau$ and $\epsilon = $ 1 or 0 for timelike or null geodesics. The use of $dx^i/dt$ and $du_i/dt$ are coordinate dependent, but within a fully defined spacetime and coordinate system, these values take on a specific meaning. To illustrate this let us consider the Alcubierre metric \cite{1994CQGra..11L..73A} given by:
\begin{equation}
    ds^2 = -d t^2+\left(d x-v_s f\left(r_s\right) d t\right)^2+d y^2+d z^2
\end{equation}
where $r_s= \sqrt{(x-x_s(t))^2 + y^2 + z^2}$. $x_s(t)$ and $v_s(t)$ are the center position and speed of the warp drive, respectively, and $f(r_s)$ is a shape function that defines the warp bubble extent from the center. Alcubierre's solution is a shift vector addition to an otherwise Minkowski spacetime. Thus, it is clear that all the characteristic warp drive features of this spacetime are sourced purely by the shift vector. We can understand this from a different perspective, using the steps and geodesic equations  \eqref{eq:geodesicEvolution} from above. Inside the passenger volume region (defined as $r_s < R$) the spacetime is constrained to be flat, meaning that $\partial_i g_{\mu\nu} = 0$ and hence $du_i/dt = 0$. The passenger geodesic motion in the coordinate system within that region is:
\begin{equation}
    \frac{dx^i}{dt} = v_s = \delta^{ij}\frac{u_j}{u^0} - \beta^i
\end{equation}
As seen from the equation above, the geodesic transport of passengers depends on their $u_i$ at the beginning and end of the transport. Alcubierre's solution allows for acceleration using a time-varying $v_s(t)$. If we consider the starting condition where observers are initially at rest ($u^i = u_i = 0$) at some point and then imagine a warp bubble forming around the passengers with the same constraints ($du_i/dt = 0$ for $r_s < R$), then $u_i(t) = 0$ at any time $t$ and the transportation of passengers within the bubble in this scenario is given by:

\begin{equation}
    \frac{dx^i}{dt} = v_s(t) = -\beta^i(t)
\end{equation}

\begin{figure}[h]
\begin{center}
\includegraphics[width = \textwidth]{"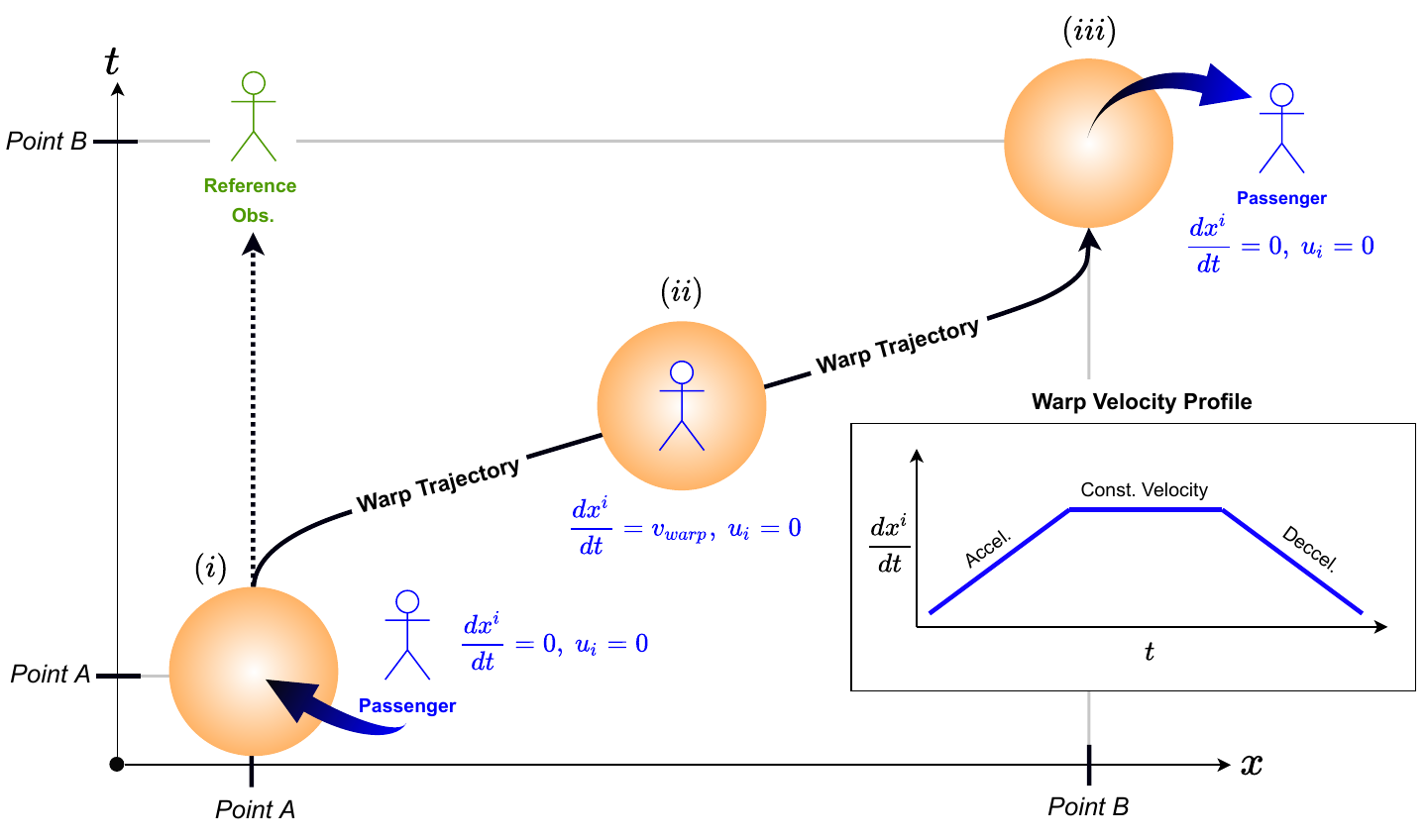"}
\caption 
{ \label{fig:exampleWarpFlight}
Example of an Alcubierre warp trajectory with three phases of flight: \\
\vspace{0.2cm} \\
\noindent \textit{(i)} Passenger enters the warp bubble at rest w.r.t to the reference observer at point A. The passenger will not have any coordinate velocity compared to the reference observer ($dx^i/dt = u_i = 0$) 
\vspace{0.4cm} \\
\noindent \textit{(ii)} Warp bubble begins its travel by accelerating up to a constant velocity. The passenger inside is geodesically transported along up to the coordinate velocity of the warp drive, but still with zero velocity as measured in its proper time ($dx^i/dt = v_{warp}$).
\vspace{0.2cm} \\
\noindent \textit{(iii)} Warp bubble decelerates to a stop at point B at rest w.r.t to the reference observer and the passenger exits the drive.
} 
\end{center}
\end{figure}

In this context, we can consider Alcubierre's warp drive solution as capable of transporting observers initially at rest with respect to an external stationary observer and up to a relative velocity of $v_s$, all accomplished using a localized shift vector in the spacetime with a flat interior region. This example of warp travel is illustrated in Figure \ref{fig:exampleWarpFlight}.

In this paper, we will focus on analyzing the constant velocity phase of warp flight. This constant velocity focus presents some challenges, as without acceleration, the boundary conditions of passengers entering the drive are less interesting in the simple comoving case. For example, if the warp drive is always at a constant velocity with its passengers having always existed comoving inside the warp drive, then no shift vector is needed since the passengers $u_i/u^0 = v_s$ by definition. Bobrick and Martire discuss this example of a constant velocity case as the basis of a physical warp solution in \cite{2021CQGra..38j5009B}, which results in a regular matter shell. In the Alcubierre example, we can gain an important context for its constant velocity phase when connected to a prior accelerating phase where we define how passengers enter the drive. If the drive undergoes acceleration with a condition of $du_i/dt = 0$, then having a shift vector during the constant velocity phase is required as the passengers prior to this point had $u_i/u^0 < v_s$ and only a shift vector can provide the required $dx^i/dt$ under the curvature constraints applied in the passenger volume.

Almost all of the warp solutions proposed in the literature are essentially variations of Alcubierre's solution and rely on the shift vector to provide the transportation of passengers in the same sense as discussed here. Van Den Broeck's \cite{1999CQGra..16.3973V} addition of a spatial term is only used to make more efficient use of energy density by volume expansion, but the shift still plays the same role in driving transportation as in the Alcubierre solution, as the spatial term is flat in the passenger volume\footnote{The non-unitary spatial term will modify the geodesic velocity as $\gamma^{ij}\beta_{j}$}. Lentz \cite{2022arXiv220100652L, 2020arXiv200607125L}, Fell and Heisenberg \cite{2021CQGra..38o5020F} and the more general Natario class \cite{2002CQGra..19.1157N}  warp metrics all use several shift vector components but follow in essence the same dynamics as described here. In the spherical symmetric metric \cite{2021CQGra..38j5009B}, no shift vector is used, but again, this is a solution of constant velocity, and shift vector addition could be required when expanding it to a more general accelerating solution. 

While shift vectors have been used extensively in the literature, a shift vector is not the only ingredient we can use to build warp drives. Spatial gradients in the lapse and metric spatial terms can affect $du_i/dt$ and then $dx^i/dt$ but require careful management of their spatial derivatives to avoid energy or tidal forces existing inside the passenger volume. The use of a shift is, in many ways, the easiest method to add geodesic transportation as it directly provides a $dx^i/dt$ term to observers based on its magnitude alone. In this work, we will focus on a warp solution that uses a shift vector to obtain the same warp drive properties as those of the Alcubierre solution but in a manner that can maintain physicality.

\subsection{The Problem of Physicality}
The condition of a physical warp drive is discussed in detail in \cite{2021CQGra..38j5009B} and our recent paper \cite{NumericalWarp2023}, but in essence, the core requirement is to satisfy the energy conditions. The Natario-class of solutions (defined by shift vectors with unit lapse $\alpha=1$ and flat spatial metric $\gamma_{ij}=\delta_{ij}$) has been shown to always violate the energy conditions \cite{2022PhRvD.105f4038S}. One possible reason for this violation is that the metrics as constructed lack the gravitational effect of regular matter. The asymptotic gravitational field produced by a gravitating spherically symmetric object is given by the Schwarzchild metric, which has an asymptotic $1/r$ dependency in the lapse and spatial terms in Schwarzchild coordinates at infinity. For general warp metrics with a \textit{compact support region}, meaning a metric whose components are bounded within some finite region that transitions to Minkowski space faster than 1/r, behaves in ways different from regular matter. This can be expressed in another way using the definition of ADM mass \cite{2007GReGr..39..521B}, which is a quantity describing the concept of mass as seen in faraway regions. Alcubierre metric and similar solutions have $M_{ADM} = 0$, as opposed to Schwarzschild metric, which for an equivalent energy density magnitude would have $M_{ADM} > 0$. However, even if the metric has non-zero ADM mass, energy condition violations can easily occur. This issue is demonstrated in the recent work of Schuster et al. in the transported Schwarzschild Drive \cite{2022arXiv220515950S}. Further still, even if the solution asymptotically approaches that of a positive matter spacetime with a positive ADM mass, in the non-vacuum warp bubble additional constraints must be applied. As a rule of thumb, the Eulerian momentum flux and pressures should be less than the energy density to satisfy the energy conditions \cite{NumericalWarp2023}. Finally, from \cite{2021CQGra..38j5009B} subluminal motion is likely another important requirement for the metric to be physical. In summary, the likely key ingredients to a physical warp drive solution can be simply stated as:
\begin{enumerate}
    \item The asymptotically flat spacetime should have a positive ADM mass.
    \item Generally, much larger positive energy density than both pressure and momentum flux in the non-vacuum warp bubble, as measured by Eulerian observers.
    \item Subluminal speeds
\end{enumerate}
These physical ingredients will be the guiding focus of the solution constructed in this paper.

\subsection{Paper Structure}
The paper is structured by first introducing the approach to building a numerical model of a warp drive in Section \ref{sec:methods}. Then, in Sections \ref{sec:shell} - \ref{sec:constantvelocitywarp}, we develop the solutions for a matter shell and its transformation to a Warp Shell through the addition of a shift vector. In Section \ref{sec:discussion} we discuss the implications of this solution and compare it to prior warp metrics. Finally, we conclude and remark on future steps in Section \ref{sec:conclusion}.


\section{Methods}\label{sec:methods}
To overcome the issues encountered by warp solutions in the past, we will use a new approach to constructing warp solutions that maintain a Schwarzschild vacuum solution at large distances with a compact stress-energy tensor. This is accomplished by adding a shift vector distribution on top of a regular shell of matter. The added shift vector is kept below a threshold that causes energy condition violation from the added momentum flux that accompanies its addition. Adding a shift vector will have a similar effect on passenger transport to that in the Alcubierre drive without any energy condition violations.

\subsection{Building the Bubble}
To find a physical solution, we utilize a moving matter shell as the foundation metric for our warp drive. This solution features a flat interior with an asymptotically-flat Schwarzchild solution outside the shell. The shell solution will be constructed in comoving coordinates in which the metric tensor does not depend on time. In this section, we provide a top-level summary of the process, while the details are found in the next section.

\begin{figure}[h]
\begin{center}
\includegraphics[width = \textwidth]{"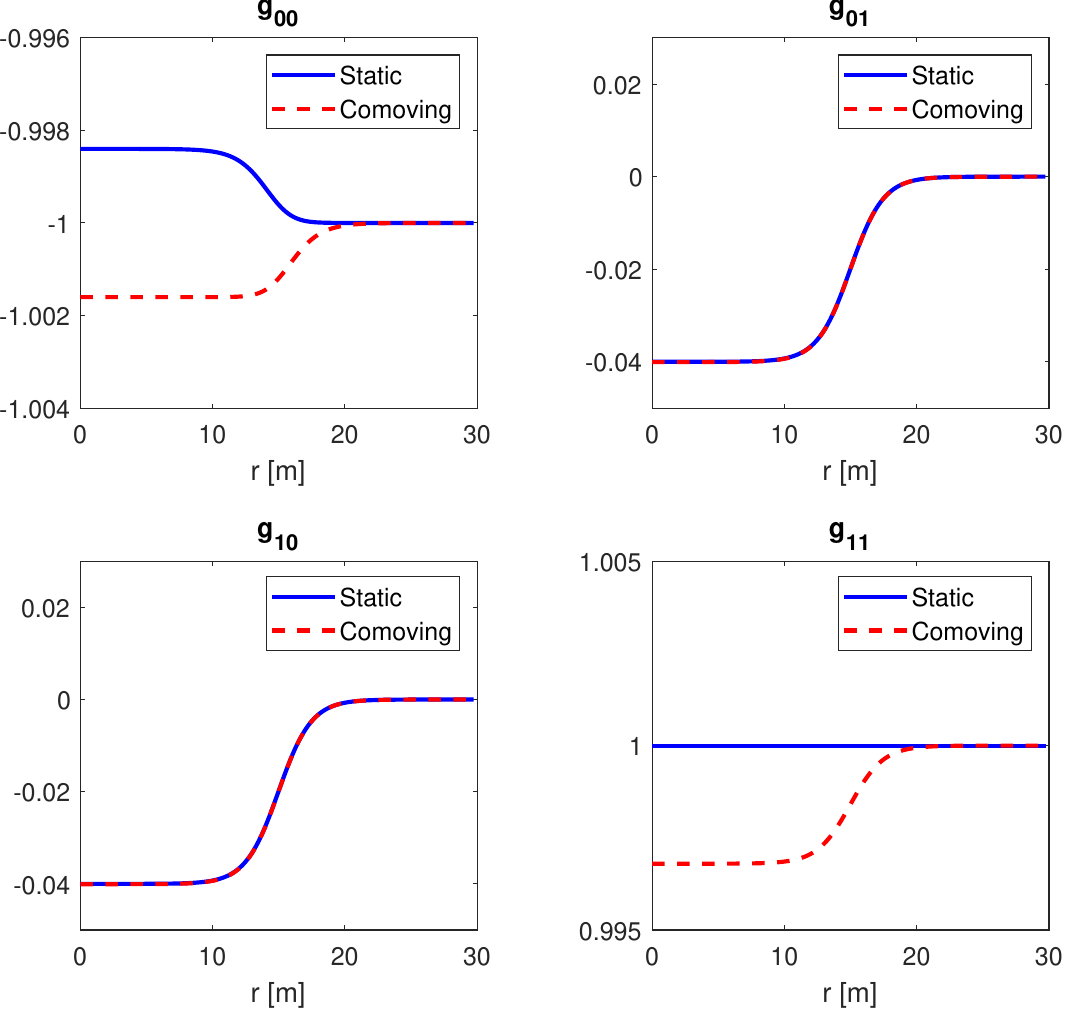"}
\caption 
{ \label{fig:LorentzAlcubierre}
Example of an Alcubierre solution transformed to a comoving frame using a Lorentz transformation. The static (blue) is the metric for the form defined by Alcubierre in \cite{1994CQGra..11L..73A}. The comoving (dotted red) is the inverse Lorentz transformation applied to the static solution. Note that the shift vector remains the same but changes occur for the spatial and lapse terms. These changes will be superseded by the matter shell terms when building the physical solution in this paper.} 
\end{center}
\end{figure}

First, we need to consider what warp solutions look like in a comoving frame. Since we plan to add a single shift vector to a shell, we look at the Alcubierre solution in a comoving frame to see the form of the shift vector we want to add. To transform Alcubierre's solution to the stationary frame of an external comoving timelike observer, we can use a Lorentz transformation shown in \ref{apx:lorentz} as determined for an observer at spatial infinity. It should be noted that this transformation limits us to a subluminal regime, but that is a natural restriction for any external comoving timelike observer and is the regime that is most likely to lead to physical solutions. Performing the Lorentz transformation to the Alcubierre metric results in a shift vector that is zero at $r_s \gg R$ and a non-zero shift vector inside the passenger volume, as shown in Figure \ref{fig:LorentzAlcubierre}.

With the transformed Alcubierre solution above, we can now compare this to the same setup for a regular matter shell, which is typically defined in a comoving frame, and see that the warp effect addition can simply be expressed as adding shift vector to the shell metric. This added shift vector $\beta_i$ is applied inside the interior of the shell, where the matter exists to manage the non-vacuum physicality constraints:
\begin{equation}
    g_{warp shell} = g_{shell}+ \delta g_{warp}
\end{equation}
where $\delta g_{warp}$ is a metric only containing a shift-vector component along a single direction:
\begin{equation}
    \delta g_{warp} = \begin{pmatrix} 0 & \beta_1 & 0 & 0 \\
    \beta_1 & 0 & 0 & 0 \\
    0 & 0 & 0 & 0 \\
    0 & 0 & 0 & 0 \end{pmatrix}
\end{equation}
The details for the shell and its warp modification are described in Sections \ref{sec:shell} and \ref{sec:constantvelocitywarp}.

As we did for the Alcubierre metric we can return to the geodesic equations \eqref{eq:geodesicEvolution} to describe how this solution would impact passengers. Although we do not model the acceleration phase here, it may be considered similar to the Alcubierre case. Specifically, the passenger region of this shell interior in constant velocity case will need to be flat, so we can again assume that $du_i/dt = 0$. We will also let $u_i = 0$, just as we did for the case of Alcubierre metric, to consider possible solutions connected to some prior acceleration phase which maintained $du_i/dt = 0$. However, this time, the shell solution will not have Minkowski spatial terms and thus the coordinate motion is now given by:
\begin{equation}
    \frac{d x^i}{d t}= - \beta^i = - \gamma^{i j}\beta_j \\
\end{equation}
Lastly, a question occasionally discussed in warp theory is the nature of the passenger transport w.r.t to the bubble motion, especially as the shift vector decreases to zero at the boundary. In the case of a constant velocity warp drive, where the metric is not changing in the comoving frame, the bubble matter itself will always by definition ``move" aligned with the passenger volume. This is a consequence of the fact that the bubble itself is a generating function for the shift vector, determined by the metric through the Einstein field equation. This is true for all constant velocity warp solutions.

\subsection{Numerical Methods}
In this section, we present a summary of our numerical method. We perform numerical analysis using the Warp Factory toolkit presented in detail in \cite{NumericalWarp2023}. Throughout this paper, we will adopt the 3+1 formalism and always report the Eulerian stress-energy tensor and its components (pressure, momentum flux, and energy density) using the methods from Section 3 in \cite{NumericalWarp2023}. In Warp Factory, the frame transformation is done on the metric tensor locally at each point using a tetrad corresponding to Eulerian observers, which also transforms the local metric to Minkowski form. This transformation tetrad is applied to the stress-energy tensor, returning the Eulerian-measured stress-energy tensor. The Eulerian observer in this frame is defined in a standard way as:
\begin{equation}
    n_{\mu} = (1, 0, 0, 0)
\end{equation}
The observation of energy density $\rho$, momentum flow $p_i$, isotropic pressure $P_i$ and stress tensor $\sigma_{ij}$ from the Eulerian tensor $T^{\hat{\mu}\hat{\nu}}$ by the Eulerain observer at any location is:
\begin{equation}
\begin{split}
    \rho =  T^{\hat{0}\hat{0}} \\
    p_i = T^{\hat{0}\hat{i}} \\
    P_i = T^{\hat{i}\hat{i}} \\
    \sigma_{ij} = T^{\hat{i}\hat{j}}
    \end{split}
\end{equation}
The process for defining and evaluating the energy conditions, which is described in detail in Section 3 of \cite{NumericalWarp2023}, is shown here for clarity. The Null Energy Condition (NEC) is given by the contraction of null observers with the stress-energy tensor at all points of the spacetime ($X$):
\begin{equation}
    \Xi_{NEC}(X) = T_{\hat{\mu}\hat{\nu}}(X) k^{\hat{\mu}} k^{\hat{\nu}} \ge 0 \ \ \forall \ \  k^{\hat{\mu}}
\end{equation}
where $k^{\hat{\mu}}$ are null observers. The Weak Energy Condition (WEC) is similar to the NEC but with the contraction of timelike observers at all points of spacetime:
\begin{equation}
    \Xi_{WEC}(X) = T_{\hat{\mu}\hat{\nu}}(X) V^{\hat{\mu}} V^{\hat{\nu}} \ge 0 \ \ \forall \ \  V^{\hat{\mu}}
\end{equation}
where $V^{\hat{\mu}}$ are timelike observers. The Strong Energy Condition (SEC) is also found using timelike observers contracted with the stress-energy tensor:
\begin{equation}
    \Xi_{SEC}(X) = \left(T_{\hat{\mu}\hat{\nu}}(X) - \frac{1}{2} T(X) \eta_{\hat{\mu}\hat{\nu}} \right)V^{\hat{\mu}} V^{\hat{\nu}} \ge 0 \ \ \forall \ \  V^{\hat{\mu}}
\end{equation}
Finally, the Dominant Energy Condition (DEC) is given by contracting the stress-energy tensor in the mixed form using the timelike observers:
\begin{equation}
    \Upsilon^{\hat{\mu}}(X) = -T^{\hat{\mu}}_{\ \ \hat{\nu}}(X) V^{\hat{\nu}}
\end{equation}
where $\Upsilon^{\hat{\mu}}\left(X\right)$ must be future pointing, meaning $\Upsilon^{\hat{\mu}}$ is either timelike or null satisfying\footnote{In this work, we flip the sign of this condition in Warp Factory so that negative values mean violations in all of the energy conditions shown.}:
\begin{equation}
    \xi_D(X) = \eta_{\hat{\mu}\hat{\nu}} \Upsilon^{\hat{\mu}}(X) \Upsilon^{\hat{\nu}}(X) \le 0 \ \ \forall \ \  V^{\hat{\mu}}
\end{equation}
The observer vector field is sampled with a spatial orientation density of 100 samples and for timelike observers and an additional velocity magnitude density of 10 samples (see \cite{NumericalWarp2023} for a detailed discussion of this method). 

\section{Shell Metric}\label{sec:shell}
The base for the warp solution is a stable matter shell. We start by constructing this shell in a comoving frame in  Schwarzschild coordinates. 

\subsection{Metric Definition}\label{sec:shellmetricdef}
\begin{figure}[ht]
\begin{center}
\includegraphics[width = \textwidth]{"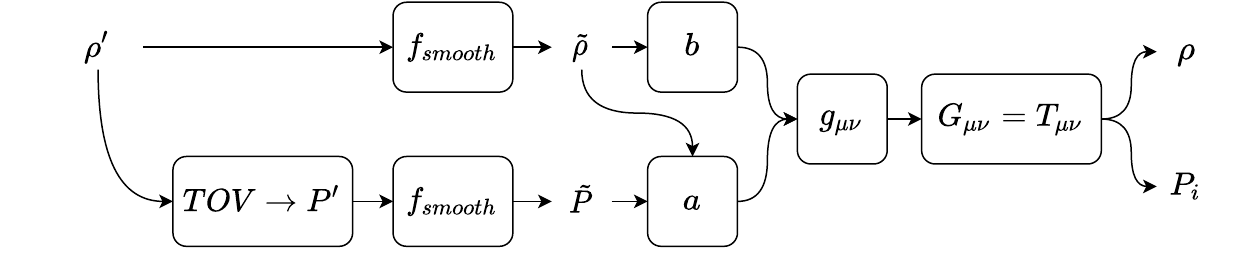"}
\caption 
{ \label{fig:shellconstructiondiagram}
Metric creation method where trial solutions are used and then modified to construct a physical shell solution. The process starts with density on the left and then generates a solution on the right.} 
\end{center}
\end{figure}

The shell solution is built starting from a general static, spherically symmetric metric, which has the form of  \cite{2004sgig.book.....C}:
\begin{equation}
    ds^2 = -e^{2a}dt^2 + e^{2b}dr^2 + d\Omega^2
\end{equation}
The functions of $a$ and $b$ can be solved using the field equation with a known stress-energy tensor. For a simple solution based on the stress-energy tensor for an isotropic fluid, this is a straightforward process where the stress-energy tensor components in the Eulerian frame are given as:
\begin{equation}
    T^{iso}_{\hat{\mu}\hat{\nu}} = \textrm{diag}(\rho,P,P,P) 
\end{equation}
However, for a stable shell the pressures $P$ can not be assumed as isotropic since the interior radius must withstand the gravity inward pressure, resulting in non-uniform pressure terms along the $\theta$ and $\phi$ directions, akin to hoop stress in a cylinder. With non-isotropic pressure, the solution takes the form:
\begin{equation}
    T^{shell}_{\hat{\mu}\hat{\nu}} = \textrm{diag}(\rho,P_1,P_2,P_3) 
\end{equation}
To solve for the non-isotropic shell solution, we will take an iterative approach to find $a$ and $b$ as modifications from the isotropic solution by changing the assumed pressure and density used to determine $a$ and $b$ in the isotropic case. A short summary of the process will be as follows:
\begin{enumerate}
\item Start with an initial guess solution for the shell metric assuming a constant density $\rho^\prime$ between the inner radius $R_1$ and the outer radius $R_2$.
\item Solve for the initial guess pressure profile $P^\prime(r)$ by assuming the pressure in the shell is isotropic and zero at $r = R_2$ using the Tolman-Oppenheimer-Volkoff (TOV) equation.  After solving the differential equation with the single boundary condition at $r = R_2$, we are left with a constant pressure inside, this pressure is set to zero for $r < R_1$ to enforce a vacuum interior.
\item The constant density assumption creates sharp boundaries at $R_1$ and $R_2$ and the isotropic pressure assumption is not valid for the $R_1$ boundary to maintain a stable shell. To address this issue, we soften the boundary using radial smoothing applied to $\rho^{\prime}$ and $P^{\prime}$ using an $f_{smooth}$ function.  
\item The smoothed $\tilde{\rho}$ and $\tilde{P}$ are then used to solve for the terms of $a$ and $b$ which build the actual metric. Solving the stress-energy tensor using the Einstein field equations then provides the true $\rho$ and $P_i$ that correspond to the metric obtained in this step.
\item Finally, the smoothing function for pressure and density is iterated upon in steps (iii) and (iv) until we find a metric that satisfies the energy conditions at the boundaries.
\end{enumerate}
The process flow for constructing the metric solution using the process described above is shown in Figure \ref{fig:shellconstructiondiagram}.

The detailed version of the process outlined above is as follows. The starting assumption of the density profile $\rho^\prime$ is that of a spherical shell with an inner radius of $R_1$ and outer radius of $R_2$ with a constant density and total mass $M$. This defines the density as:
\begin{equation}\label{eq:rho}
    \rho^\prime(r) = 
    \begin{cases}
     0 & 0 \leq r \leq R_1 \\
     \frac{3}{4\pi}\frac{M}{R_2^3-R_1^3} & R_1 \leq r \leq R_2 \\
     0 & R_2 \leq r < \infty
    \end{cases}
\end{equation}
and the associated cumulative mass profile $m^\prime(r)$ is just the integration of the density below a given radius $r$, which results in:
\begin{equation}
    m^\prime(r) = \int_0^r 4 \pi r^2 \rho^\prime(r) dr =
    \begin{cases}
     0 & 0 \leq r \leq R_1 \\
    M\left(\frac{r^3-R_1^3}{R_2^3-R_1^3}\right) & R_1 \leq r \leq R_2 \\
     M & R_2 \leq r < \infty
    \end{cases}
\end{equation}

From the density and cumulative mass definitions, we can numerically solve the TOV equation for $P^\prime$ when $R_1<r<R_2$ with a boundary of zero pressure at $r = R_2$ and enforce that $P^\prime = 0$ for $r < R_1$:
\begin{equation}\label{eq:TOV}
    \frac{dP^\prime}{dr} = 
    \begin{cases}
     0 & 0 \leq r < R_1 \\
    -G\left(\rho^\prime/c^2+P^\prime/c^4\right)\left(m^\prime/r^2+4\pi r P^\prime/c^2\right)\left(1-\frac{2Gm^\prime}{c^2 r}\right)^{-1} & R_1 < r \leq R_2 \\
    0 & R_2 \leq r < \infty
    \end{cases}
\end{equation}

\begin{figure}[ht]
\centering
    \includegraphics[width = \textwidth]{"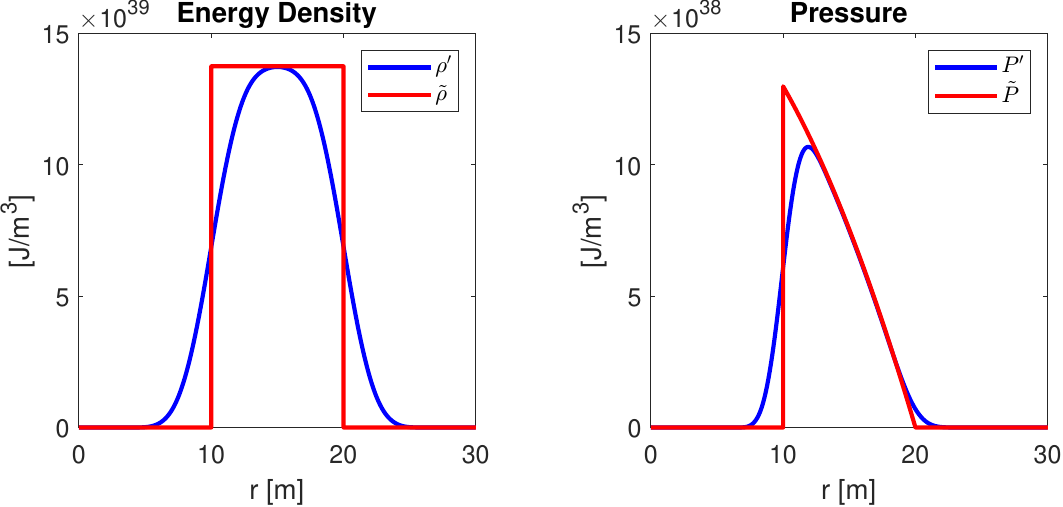"}
    \caption{Density and pressure profiles before and after 
 smoothing for constructing the Shell metric.}\label{fig:ShellSmoothing}
\end{figure}

\begin{figure}[ht]
\centering
    \includegraphics[width = \textwidth]{"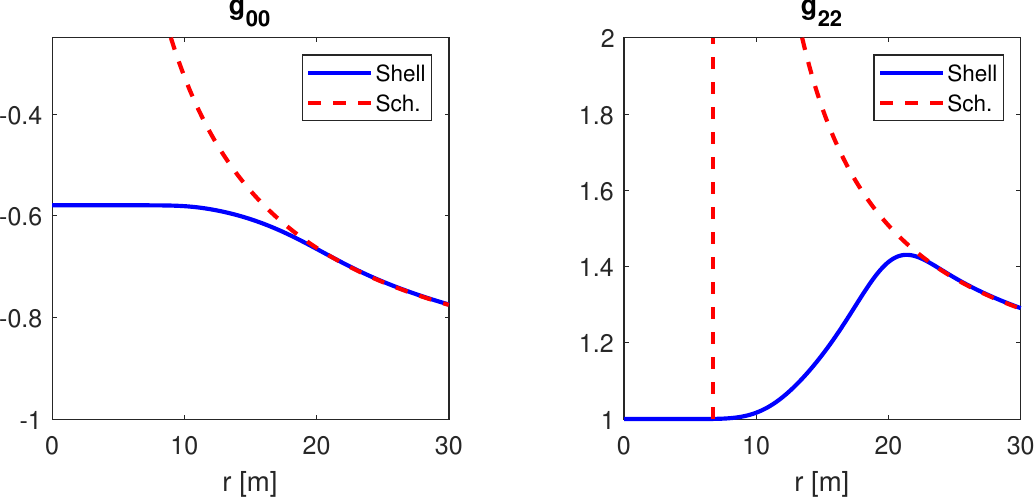"}
    \caption{Shell and Schwarzschild metric components for a slice along the y-axis. Only the non-Minkowski components for this slice are shown. In $g_{22}$ the vertical dashed line is where $r = r_s$ for the reference Schwarzschild metric, when the sign flips for the spatial parts. }\label{fig:ShellMetric}
\end{figure}

This initial solution will have issues at $R_1$ and $R_2$ due to the discontinuity of the density and pressure, this problem is alleviated by applying a numerical smoothing to both $\rho^{\prime}$ and $P^{\prime}$:
\begin{equation}\label{eq:inputPandrho}
    \begin{split}
        &\tilde{\rho} = f_{smooth}(\rho^\prime) \\
        &\tilde{P} = f_{smooth}(P^\prime) \\
    \end{split}
\end{equation}
The smoothing function applied uses a moving average, which is a lowpass filter with filter coefficients equal to the reciprocal of the span of the average \footnote{See MATLAB `smooth' function for more details}. The smoothing itself will fix the discontinuity by having finite values of derivatives at the boundaries while maintaining a physical solution. Selecting the smoothing function coefficients is found iteratively until the solution has no violations of the Null, Weak, Dominant, and Strong energy conditions. Once the smoothing is applied, we must recompute the new mass profile as before with the new density $\tilde{\rho}(r)$:
\begin{equation}
    m(r) = \int_0^r 4 \pi r^2 \tilde{\rho}(r) dr
\end{equation}
The smoothed values of pressure and density can now be used to solve the metric terms of $a$ and $b$. The mass profile directly determines $b$ which provides $e^{2b}$ as a simple extension of the Schwarzschild solution where $M = m$ \cite[Section 5.8, Eq. 5.143]{2004sgig.book.....C}:
\begin{equation}
    e^{2b} = \left(1 - \frac{2Gm}{c^2 r} \right)^{-1}
\end{equation}
The second term of $e^{2a}$ is found by solving for $a$ \cite[Section 5.8, Eq. 5.152]{2004sgig.book.....C}:
\begin{equation}\label{eq:solveFora}
    \frac{d a}{dr} = G\left(\frac{m}{c^2 r^2} +\frac{4\pi r}{c^4} \tilde{P}\right)\left(1-\frac{2Gm}{c^2 r}\right)^{-1}
\end{equation}
This equation is integrated using the condition that at $r \gg R_2$ the boundary is set by $e^{2a} = e^{-2b}$, which corresponds to a Schwarzchild solution in the vacuum region.  The span, or window, $s$ of the smoothing function is selected differently between density and pressure, where we have found that a ratio between the span of density and pressure $s_\rho / s_P \approx 1.72$ works \footnote{The specific numbers in the setup were found by trial and error. Improved approaches could use more complex techniques than simple moving average smoothing to resolve the boundary violation issues.}. The moving average smoothing is applied four times, with the same span and ratios, to the density and pressure for the final solution. For a shell with parameters: $R_1 = 10$ m, $R_2 = 20$ m, $M = 4.49\times 10^{27}$ kg  ($2.365$ Jupiter masses)\footnote{Selection of the mass parameter is to allow the most amount of the shift vector to the drive while balancing physicality, given the selected radial distribution of the shift vector.} the pressure and density before and after smoothing are shown in Figure \ref{fig:ShellSmoothing}. The process so far provides a solution for the metric in spherical coordinates. The last step is to transform this solution to pseudo-Cartesian coordinates, which are convenient for defining the numerical grid, by changing the coordinates and the coordinate differentials using the standard spherical to Cartesian relations. When doing this numerically, the radial solutions for $e^{2a}$ and $e^{2b}$ are interpolated to the Cartesian grid points using Legendre polynomials. The metric, which is built from these parameters, is plotted in Figure \ref{fig:ShellMetric}.

\begin{figure}
\centering
    \includegraphics[width = \textwidth]{"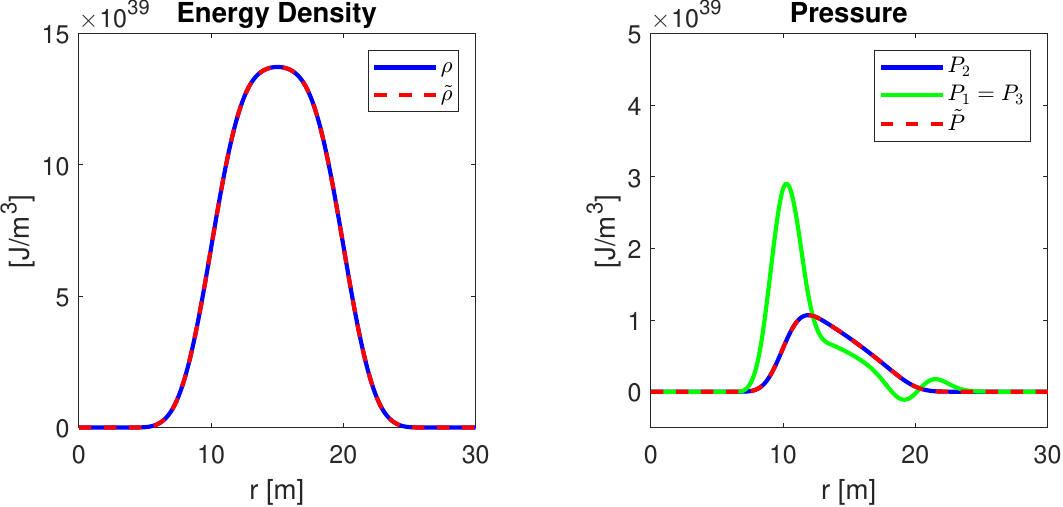"}
    \caption{Shell stress-energy components for a slice along the y-axis. Only the non-zero components are shown for this slice. The $\tilde{\rho}$ and $\tilde{P}$ lines are the smoothed density and pressure as computed from Eq. \eqref{eq:inputPandrho}. Note that the energy density is scaled by a factor of c$^2$.}\label{fig:ShellEnergy}
\end{figure}

\subsection{Stress-Energy Tensor}
\begin{figure}[ht]
\centering
    \includegraphics[width = \textwidth]{"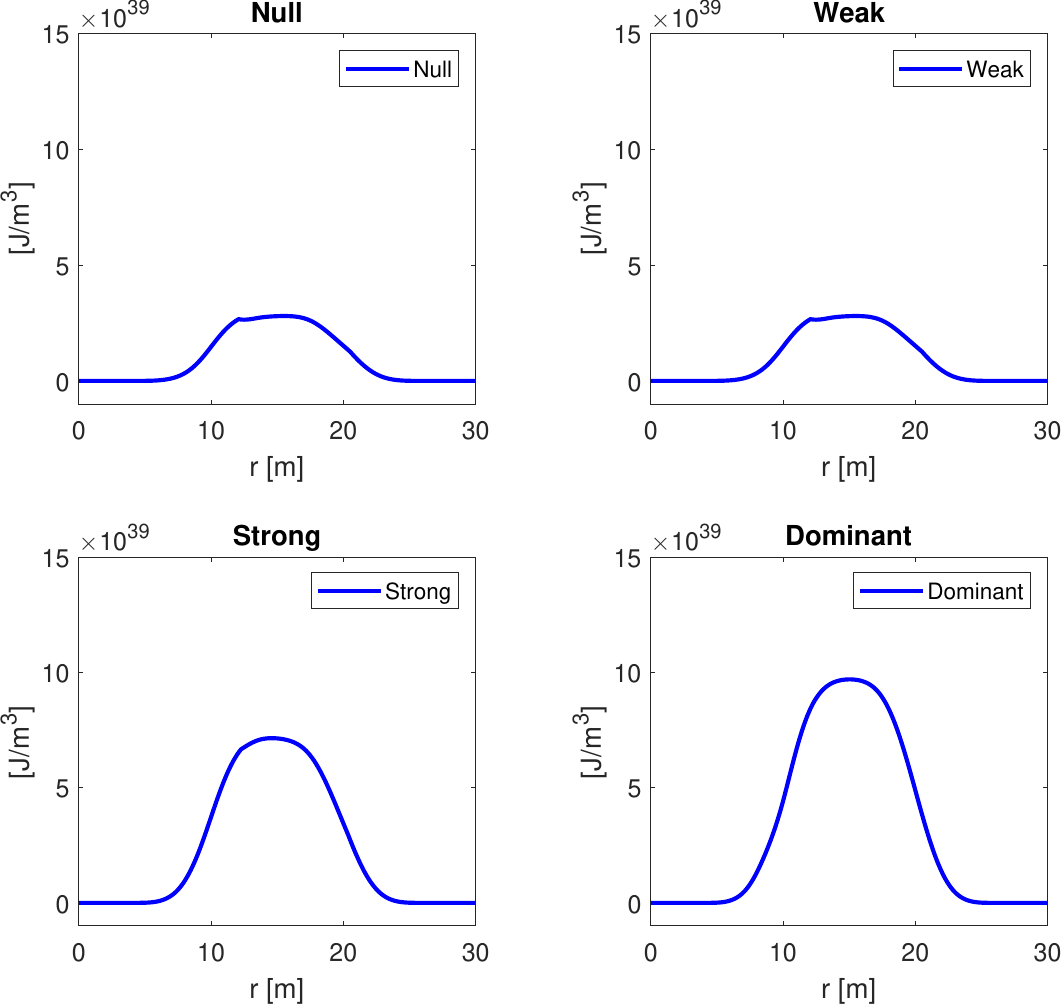"}
    \caption{Shell energy conditions for a slice along the y-axis. Negative values represent violations of the condition. No negative values are found. Units are in $[j/m^3]$}\label{fig:ShellViolation}
\end{figure}

 The resulting stress-energy terms, as measured by Eulerian observers, are plotted in Figure \ref{fig:ShellEnergy}. Along each of the principal coordinate directions, the input pressure $\tilde{P}$ used to solve for $a$ from Eq. \eqref{eq:solveFora} is equal to the calculated stress-energy pressure $P$ along that direction since it is aligned with the radial direction. The pressures along the x- and z- directions are equal and differ from the y-pressure, with a large spike on the inner bound of the shell. The choice to smooth the pressure and density is made purposely to find a solution with non-isotropic pressures, which is modified from the isotropic with a smoothing filter. For a static shell, the inner boundary at $R_1$ requires a difference in pressure between the radial pressures and the angular pressure to ensure the shell is stable from gravitational collapse. This manifests as a kind of hoop stress around the inner radius of the shell. It is also important that these pressures are all lower in magnitude than the value of the energy density at that point to ensure that the shell is physical. For realistic materials that may have a limited range of pressures possible, this requirement can always be satisfied by making the shell large and modifying the density profile, hence, reducing the gravitational forces. The physicality of the solution is demonstrated by checking the energy conditions using the Warp Factory Toolkit \cite{NumericalWarp2023}, shown in Figure \ref{fig:ShellViolation}. No energy condition violations exist beyond the numerical precision limits that exist at $10^{34}$ in this setup (see \ref{apx:erroranalysis} for a detailed discussion on errors and numerical limitations).

\clearpage
\section{Constant Velocity Warp Shell}\label{sec:constantvelocitywarp}
As described in the introduction, a warp drive that can enable the transport of different observers can do so using a shift vector inside the passenger volume. Therefore, the task is to add a shift vector field to a shell solution while maintaining the energy conditions.

\subsection{Metric Definition}
\begin{figure}[ht]
\centering
    \includegraphics[width = \textwidth]{"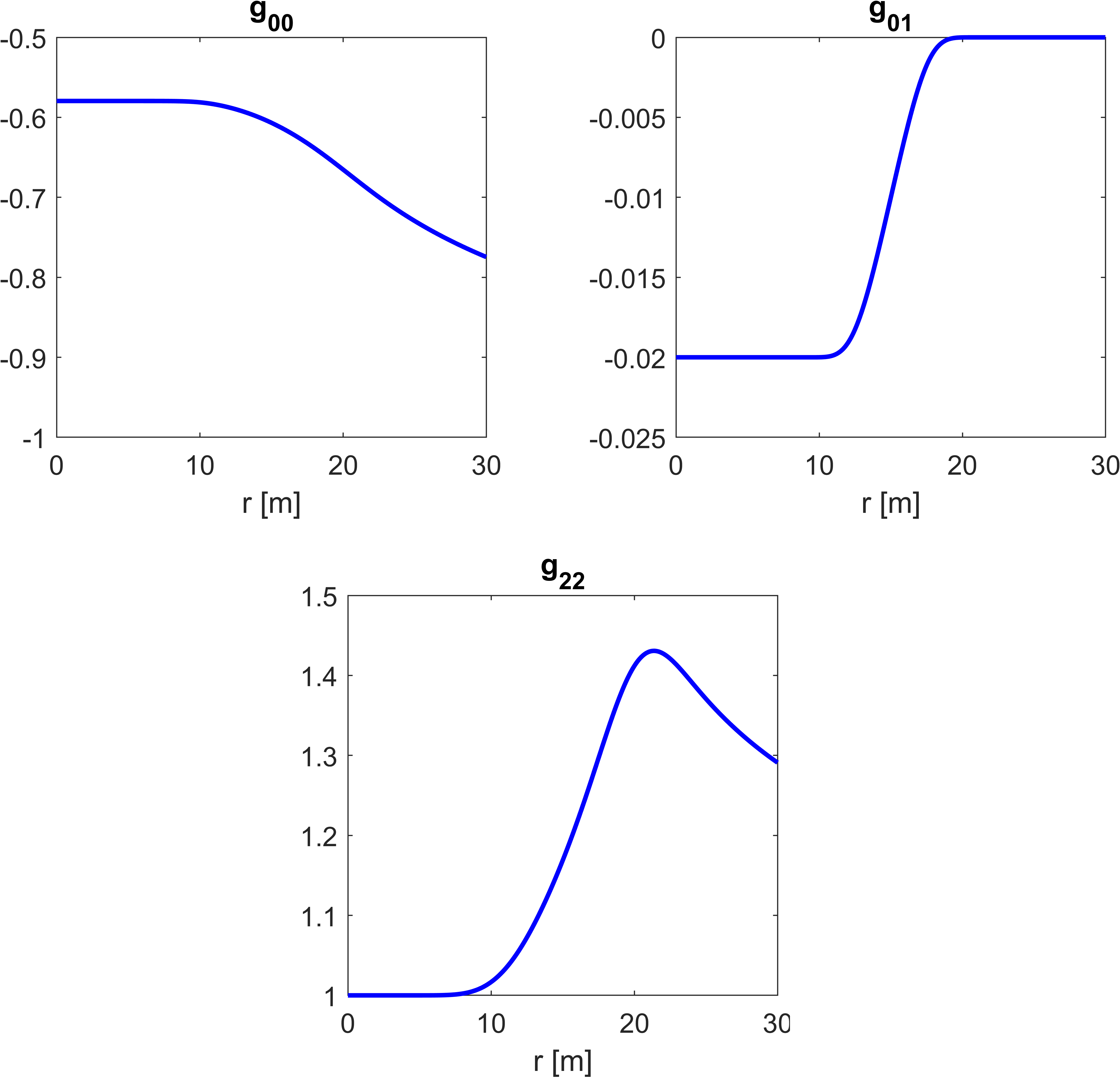"}
    \caption{Constant velocity Warp Shell metric components for a slice along the y-axis. Only the non-Minkowski components for this slice are shown. Direction of motion is along +X.}\label{fig:ShellBoostedWarpMetric}
\end{figure}

From the Shell solution constructed in Section \ref{sec:shell}, we now modify the interior region of the comoving shell to have a shift vector along the direction of motion, in this case along x. The modification must follow a few constraints to create a sensible warp drive within our definition of warp:
\begin{enumerate}
    \item The interior region should remain flat, meaning all spatial derivatives of the metric are zero ($\partial_i g_{\mu\nu} = 0$). Such a choice ensures that the passengers will be in a vacuum and experience no tidal forces.

    \item The transition region of the shift vector must occur between $R_1$ and $R_2$ and smoothly connect with the exterior solution at $R_2$ where $\beta_i = 0$. 
\end{enumerate}
The modification of shift will modify the $g_{01}$ term as:
\begin{equation}
    g^{warp}_{01} = g_{01} - S_{warp}(r)\left(g_{01} + \beta_{warp}\right)
\end{equation}
where $S_{warp}$ is a compact sigmoid function defined as:
\begin{equation}
S_{warp}(r) =
\begin{cases}
    1  & r < R_1+R_{b} \\
     1-f(r) &  R_1+R_{b} < r < R_2 - R_{b} \\
     0  & r > R_2-R_{b}
\end{cases}
\end{equation}
and $f(r)$ is given by:
\begin{equation}
    f(r) = \left(\exp\left[(R_2-R_1)\left(\frac{1}{r-R_2}+\frac{1}{r-R_1}\right)\right]+1\right)^{-1}
\end{equation}
where $R_b>0$ is a buffer region to ensure the derivatives are interior to the bubble. We construct a matter shell with the same parameters as in Section \ref{sec:shell}. Varying the values of $\beta_{warp}$, we find that the addition of shift inside the shell is possible for $\beta_{warp} = 0.02$ without any energy condition violation\footnote{This is likely not an upper limit as optimizations could be considered.}. The components for this metric are plotted in Figure \ref{fig:ShellBoostedWarpMetric}.

\subsection{Physicality}
\begin{figure}[h]
\centering
    \includegraphics[width = \textwidth]{"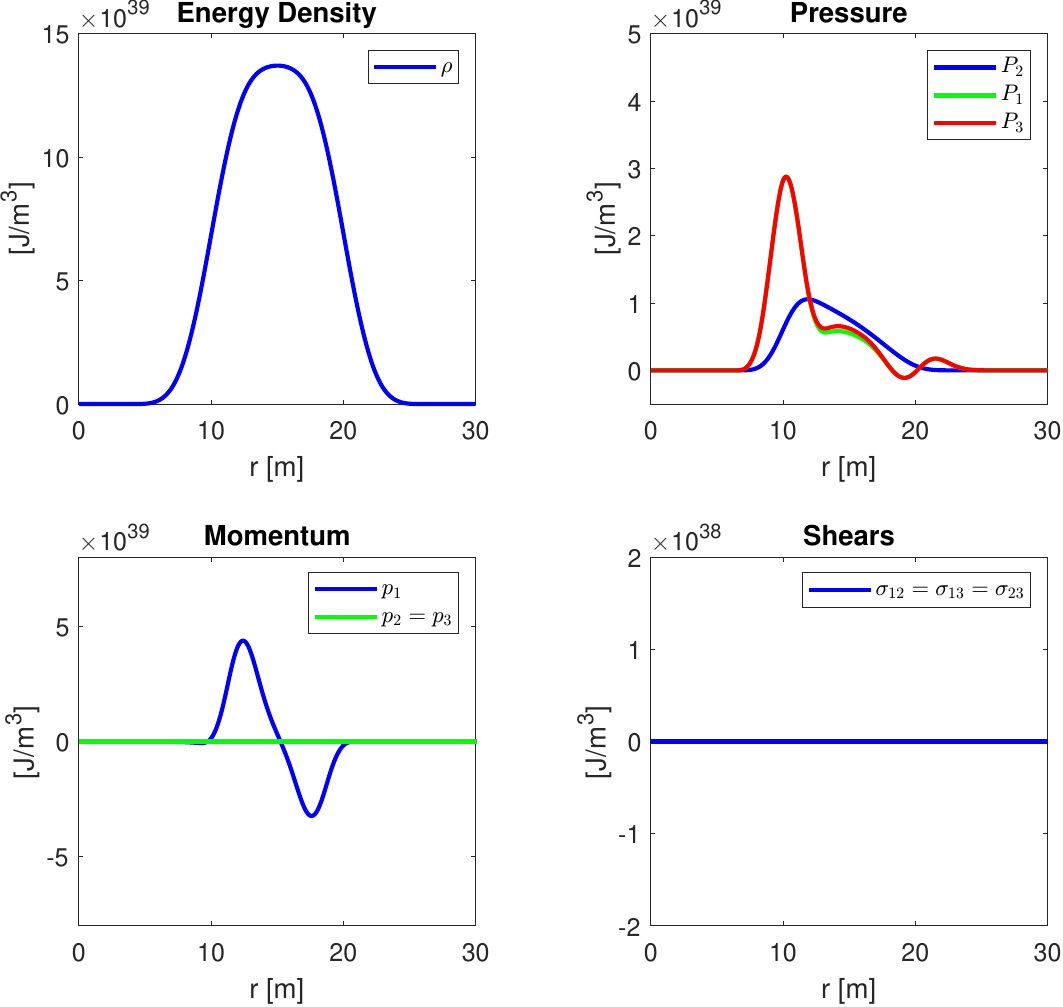"}
    \caption{Constant velocity Warp Shell stress-energy components for a slice along the cartesian y-axis. The direction of motion is along +X. Note that the energy density is scaled by a factor of c$^2$ and the momentum density by a factor of c.}\label{fig:ShellBoostedWarpEnergy}
\end{figure}

\begin{figure}[ht]
\centering
    \includegraphics[width = \textwidth]{"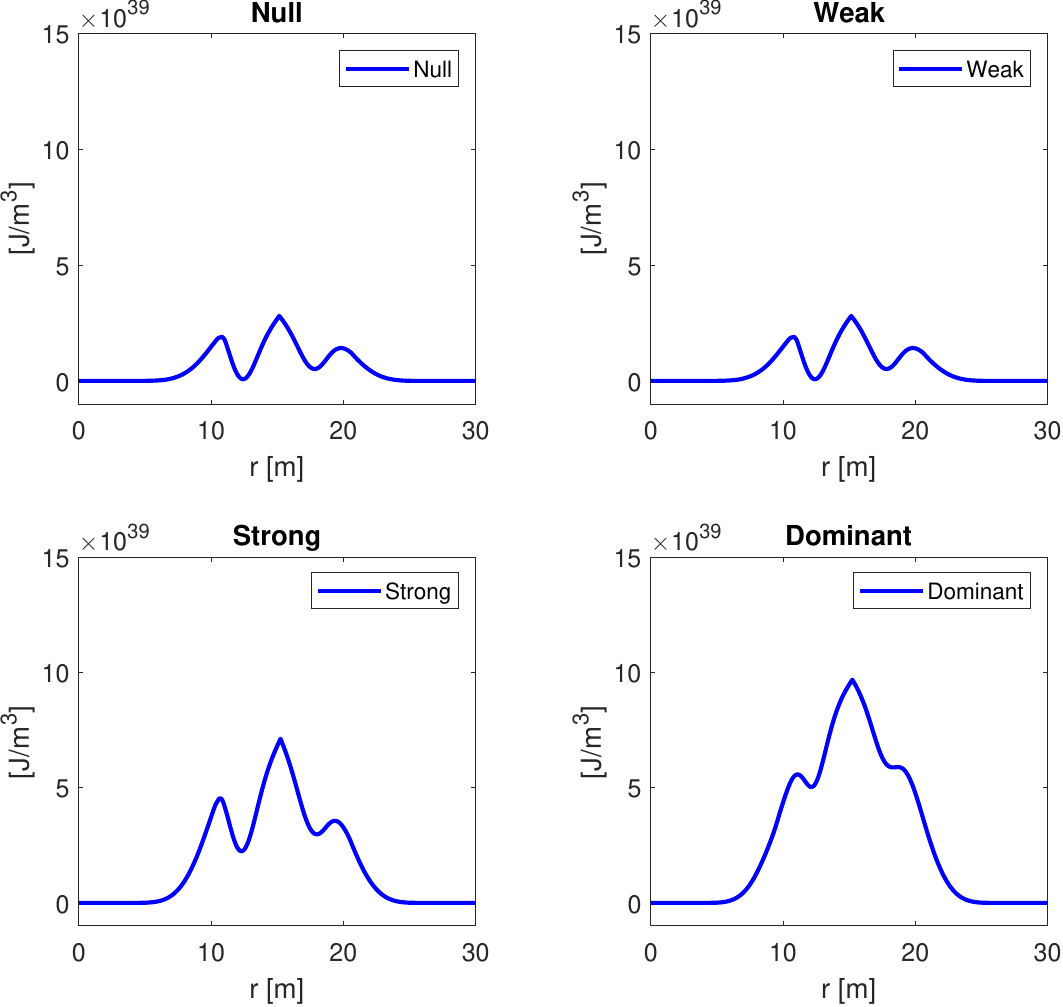"}
    \caption{Constant velocity Warp Shell energy conditions for a slice along the y-axis. The direction of motion is along +X. Negative values represent violations of the condition. No negative values are found.}\label{fig:ShellBoostedWarpViolation}
\end{figure}

 To understand the physicality of this solution, we start by plotting the resulting stress-energy terms in Figure \ref{fig:ShellBoostedWarpEnergy}. The energy density remains mostly unchanged compared to that of a standard moving shell, but the modification of the shift vector causes a difference in the momentum and pressure values for Eulerean observers. The change in the momentum is most noticeable compared to the shell metric, which had zero momentum density between $R_1$ and $R_2$. This modified solution has both positive and negative momentum density around $r \approx (R_2-R_1)/2$. This is indicative of a circulation pattern forming in the momentum flow of the shell. The same kind of momentum flow structure is also observed for an Alcubierre solution  \cite{NumericalWarp2023}. The energy conditions are evaluated for this metric and are shown in Figure \ref{fig:ShellBoostedWarpViolation}. Modification of the shift vector in this fashion has no impact on the violation compared to the normal matter shell solution.

\begin{figure}[hbt!]
\centering
\includegraphics[width=0.9\textwidth]{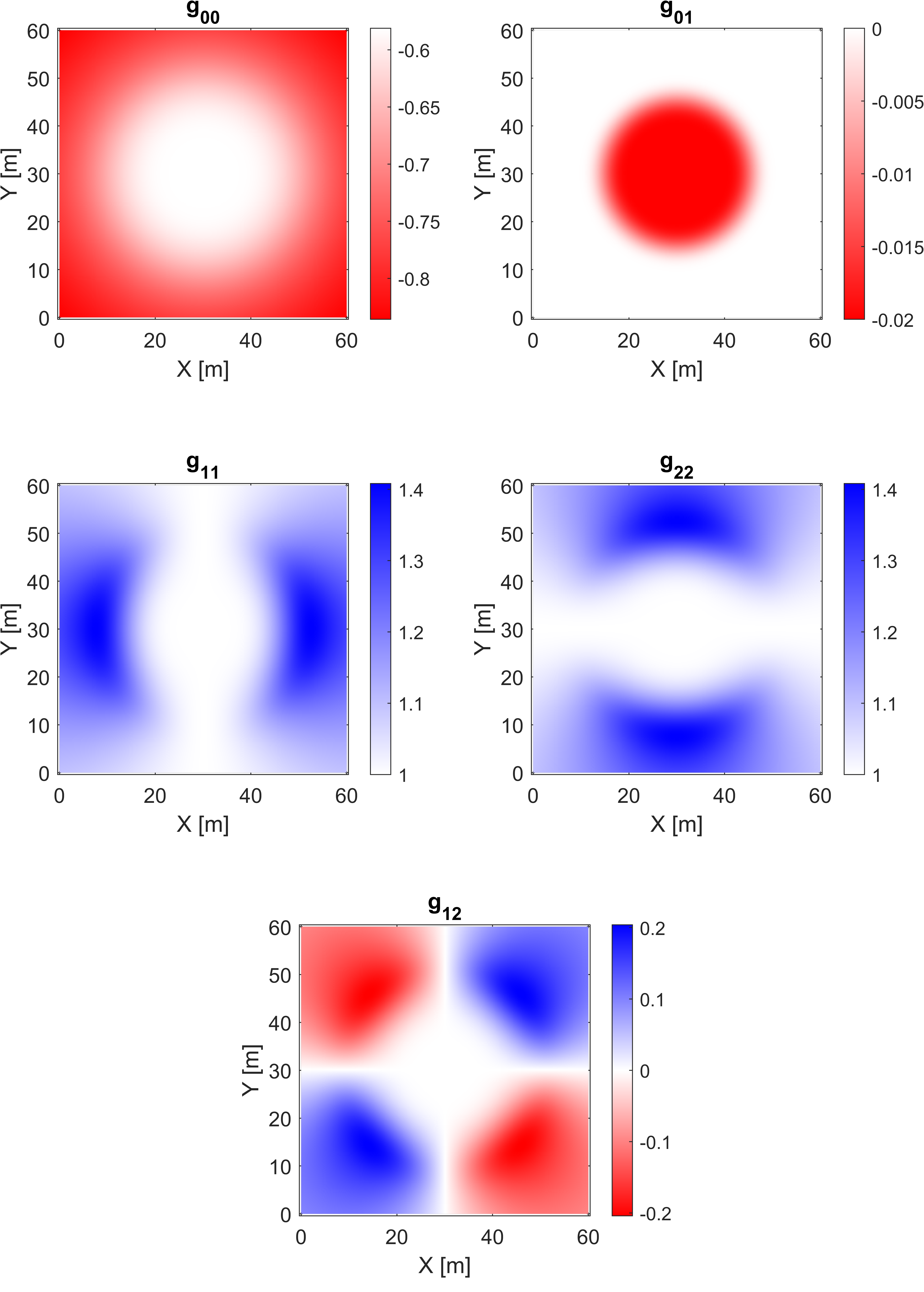}
\caption{Metric for the constant velocity Warp Shell in the comoving frame. The direction of motion is along +X. The cross-section is centered in Z. Only the non-zero cross-sections are shown.}\label{fig:WS_Metric}
\end{figure}

\begin{figure}[hbt!]
\centering
\includegraphics[width=\textwidth]{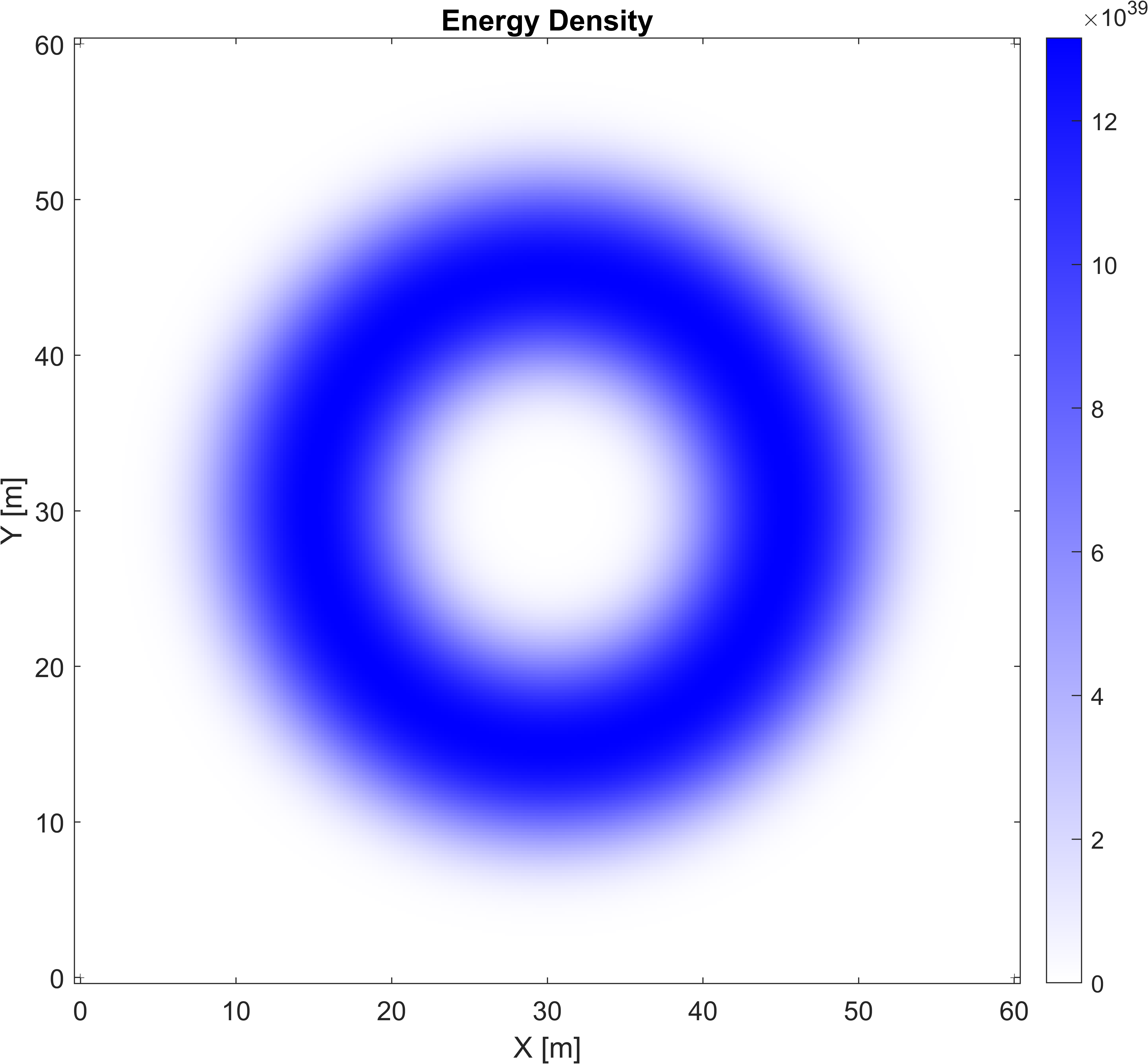}
\caption{Energy density for the constant velocity Warp Shell. The direction of motion is along +X. The cross-section along Z is aligned with the bubble center. Units are $[J/m^3]$}\label{fig:WS_Eden}
\end{figure}

\begin{figure}[hbt!]
\centering
\includegraphics[width=\textwidth]{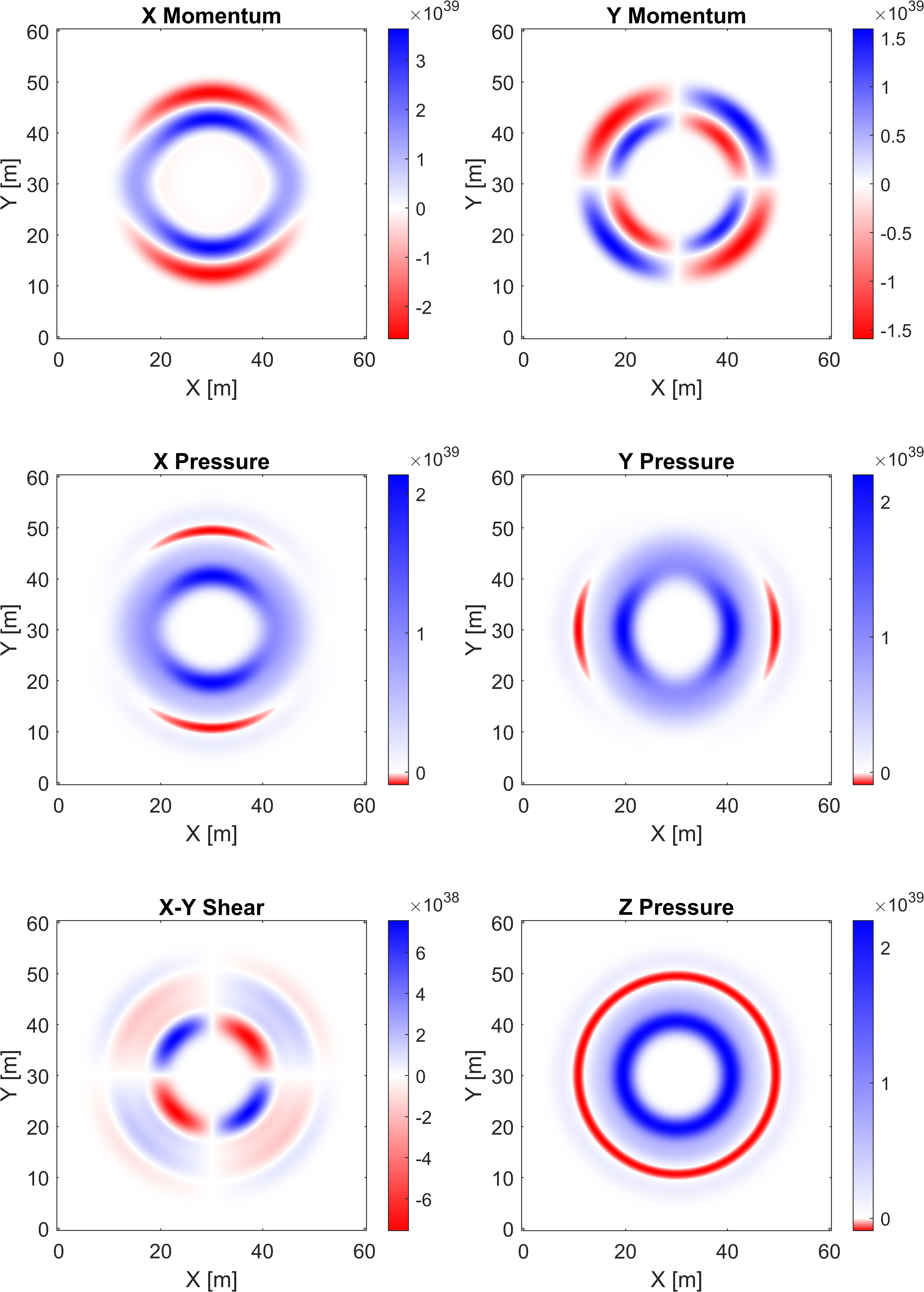}
\caption{The stress-energy tensor for the constant velocity Warp Shell in the comoving frame, for Eulerian observers. The energy density is shown in Figure \ref{fig:WS_Eden}. The direction of motion is along +X. The cross-section along Z is aligned with the bubble center. Only the non-zero cross-sections are shown. Units are $[J/m^3]$}\label{fig:WS_Etensor}
\end{figure}

\begin{figure}[hbt!]
\centering
\includegraphics[width=\textwidth]{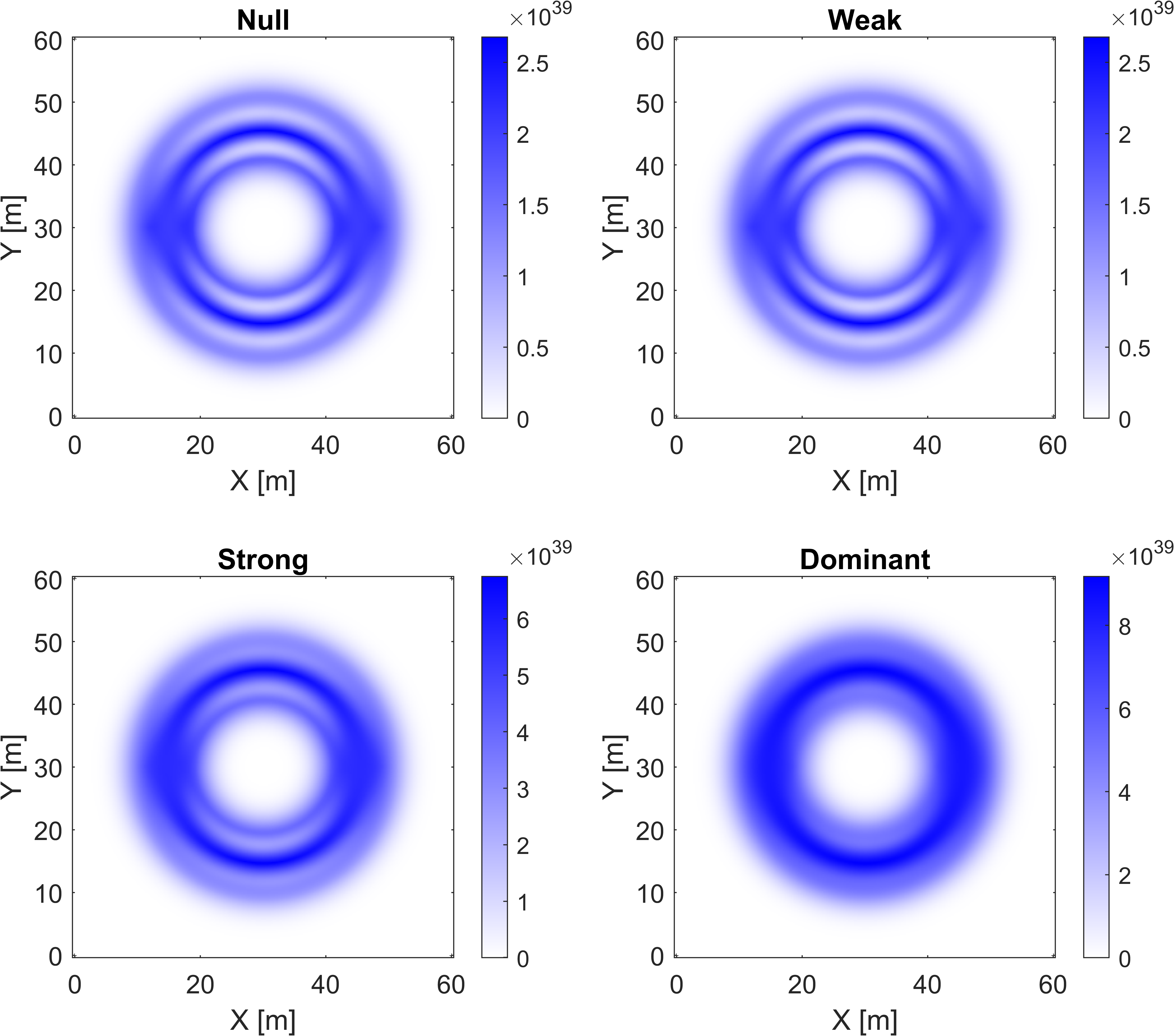}
\caption{Energy condition evaluation for the constant velocity Warp Shell. The direction of motion is along +X. The cross-section along Z is aligned with the bubble center. The minimum value across all observers is shown. Positive (blue) and zero (white) are physical and negative (red) is violating. Units are in $[J/m^3]$}\label{fig:WS_Econd}
\end{figure}

 Surf plots of the solution for a slice centered in the $Z$ direction are shown for the metric in Figure \ref{fig:WS_Metric}, energy density in Figure \ref{fig:WS_Eden}, other components of stress-energy in Figure \ref{fig:WS_Etensor}, and energy condition evaluations in Figure \ref{fig:WS_Econd}.
 
\clearpage

\subsection{Cross-Sections}

\clearpage

\section{Discussion}\label{sec:discussion}

\subsection{Measuring Shift}
\begin{figure}[ht]
\centering
\includegraphics[width=\textwidth]{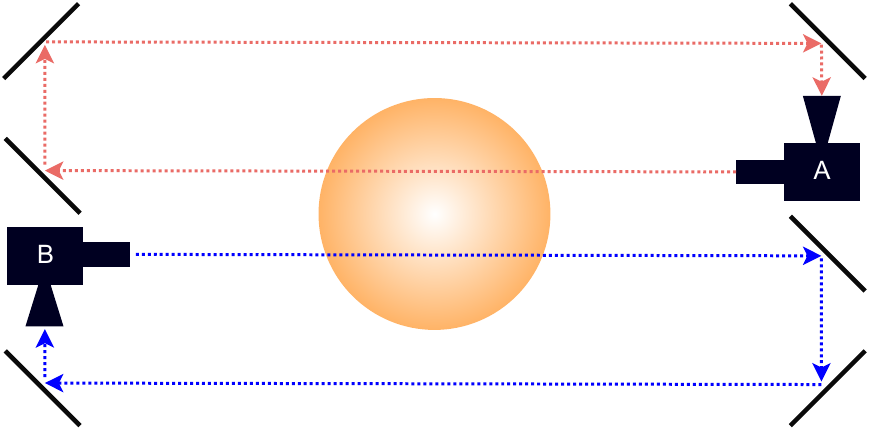}
\caption{Diagram of the light-ray test. The emitters, detectors, and mirrors are comoving with the shell of interest. Note that both beams pass through the center, but are offset in the diagram for visual clarity. Emitter-detector B is vertically aligned with the mirrors on the left and emitter-detector A is vertically aligned with the mirrors on the right. Emitter-detectors A and B are equidistant to the center of the shell. The return path of the two light beams can be anywhere outside of the shell. The Warp Shell's warp effect is in the horizontal direction away from B and toward A.}\label{fig:LightRayTest}
\end{figure}

The addition of a shift vector to the passenger volume of a shell creates several changes to the solution. To fully differentiate a warp shell from a normal matter shell, an invariant test can be constructed using a comparison of light rays traveling through the bubble, measuring the difference in transit time between two paths of rays as they transit along and against the shift vector direction. Since the metric is already defined in a comoving frame, we simply have to run null geodesics through the center of the shell directed forward and backward along the direction of the shift vector and record the proper time $\delta t$ for each photon to return as measured at the emitting points, ignoring photon interaction with the stress-energy tensor. In Figure \ref{fig:LightRayTest}, a diagram of the test setup is shown for each of the photon paths as they travel through the shell and return to the emitting point. This test configuration is constructed within Warp Factory and the light-ray times are numerically determined. Running this test we find that the Warp Shell (from Section \ref{sec:constantvelocitywarp}) has $\delta t \approx 7.6 $ ns and the Matter Shell (from Section \ref{sec:shell}) has $\delta t = 0$ ns. As expected, a normal shell has an equal transit time between both light rays, whereas the Warp Shell has a difference in transit time depending on the ray's direction through the Warp Shell. This delay is not a unique feature of the Warp Shell in this paper but is also true of other proposed warp drives that utilize a shift vector. Using Warp Factory, we conducted the same numerical experiment for a few of the warp drives discussed in the literature, and all of them have a $\delta t > 0$, shown in Table \ref{tab:metricFrameDrappingComparision}\footnote{Reference warp metrics are converted to a comoving frame using a Galilean transformation to their coordinates.}. This experiment demonstrates a Lense-Thirring effect exists for warp drives with shift vectors, creating a \textit{linear frame dragging}\footnote{An example uses photon paths circling a Kerr black hole which has a transit difference depending on traveling with or against the rotation, this same effect occurs here except these paths are straight through the warp bubble center. Both paths in either example are aligned with or against the shift vector \cite{2021Univ....7..388C}.}. Since the photon travel time is a measurable quantity, the shift-vector modification of the shell metric cannot be reduced to a coordinate transformation.
 
\begin{table}[b]
\caption{Comparison of time delay between different warp models and the matter shell for $v_{warp} = 0.04$ c.}

\centering
\vspace{0.2in}
\label{tab:metricFrameDrappingComparision}
\begin{tabular}{C{5.5cm}C{5.7cm}C{2.7cm}}
    \midrule 
    Name & Parameters & $\delta t$ [ns]\\
    \midrule 
    \textbf{Alcubierre} \cite{1994CQGra..11L..73A} & 
    $R$ = 15 m & 
    8.0 \\ 
    \noalign{\smallskip} 
    
    \textbf{Van Den Broeck} 
    \cite{1999CQGra..16.3973V} & 
     $R_1$ = 10 m, $R_2$ = 15 m, \ \ \ \ \ \ \ \ \ \ \ \ \ \ \ \ \ \ \  $\alpha$ = $0.1$ & 
     9.1 \\ 
     \noalign{\smallskip} 
    
    \textbf{Modified Time} 
    \cite{2021CQGra..38j5009B} & 
     $R$ = 15 m, A = 2 & 
     6.7 \\ 
     \noalign{\smallskip}

    \textbf{Matter Shell} \ \ \ \ \ \ \ \ \ \ \  \ \ 
    (Sections \ref{sec:shell}) & 
    $R_1$ = 10 m, $R_2$ = 20 m, \ \ \ \ \ \  \ \ \ \ \ \ \ \ \  M = $4.49 \times 10^{27}$ kg & 
    0 \\ 
    \noalign{\smallskip}
 
    \textbf{Warp Shell} \ \ \ \ \ \ \ \ \ \  \ \ \ \  
    (Section \ref{sec:constantvelocitywarp}) &
    $R_1$ = 10 m, $R_2$ = 20 m, \ \ \ \ \ \ \ \ \ \ \ \ \ \ \ \  M = $4.49\times 10^{27}$ kg & 
    7.6 \\  
    \bottomrule
\end{tabular}
\end{table}

\subsection{Positive Energy Density}
Whether or not a spacetime satisfies the energy conditions is best understood by the relationships between energy density, pressures, and momentum flux in the Eulerian frame ($T^{\hat{\mu}\hat{\nu}}$) \cite{NumericalWarp2023} since both the NEC and WEC are just expressions of how the different elements of the stress-energy tensor are perceived by different observers. In the Eulerian frame, the observer contraction can be simplified into roughly a weighting of pressure and momentum flux compared to the energy density\footnote{The full energy conditions are determined by contracting the set of all observers (null and timelike) with the tensor, which weights all of the different tensor elements together.}. We can generally say that a physical solution addressing those conditions can exist only if, for Eulerian observers, the energy density is larger than the magnitude of all of the other tensor terms combined.

One method of creating positive energy is to use spherical matter shell solutions that have a defined ADM mass \cite{2007GReGr..39..521B}. In a coordinate system that is asymptotically Minkowski, these are Schwarzschild-like solutions that can be parameterized by their ADM mass (given by $M$ in this paper). Building our warp solution from a matter shell allowed us to use the ADM mass parameter to engineer positive energy density into the solution, while the modified shift vector gave us a warp effect by creating a linear frame dragging inside the shell. However, the amount of mass is limited by the shell radius and thickness so as not to produce an event horizon within the shell ($R_{shell} > 2G M_{shell}/c^2$), so only a limited amount of energy density can be added. Increasing the shift vector will continue to add more momentum flux to the stress-energy tensor, so there is an upper limit to the magnitude of the shift vector that keeps the warp drive physical before the momentum flux exceeds energy density. This upper limit is a future direction of work. We can say that the shift vector distribution considered here is very conservative in terms of its magnitude since we keep the shell at a constant density. However, there are certainly ways to greatly improve this through optimizing the shift vector and energy density profiles to strategically place energy density where the momentum flux is highest.

Another interesting point to note is that, while there does exist a shift vector in the direction of A in the light ray setup, the Shapiro time delay \cite{1964PhRvL..13..789S}, the delay of light travel time due to gravitational time dilation, from B to A is still a delay compared to the propagation time in the corresponding flat region. This is in contrast to the Alcubierre metric, in which an advance is perceived. The presence of a changing lapse rate creates this result, a feature the Alcubierre metric does not have, that is related to the nature of the solution having ADM mass. This constraint may be another important aspect of physicality \cite{1999AIPC..493..301V}, namely that physical solutions might require a changed lapse rate which maintains a Shapiro time delay over an advance.

\subsection{Acceleration}
The warp solution created here is evaluated for the constant velocity case, but the immediate question is how it applies to the acceleration phase. One possible approach is to have the bubble accelerate by simply accelerating the coordinate center and increasing the magnitude of the shift vector accordingly. However, this approach gives the exact same issue as the Schwarzschild Drive \cite{2022arXiv220515950S}, which takes a regular black hole solution and simply moves its center through the timeslices. This approach changes the metric such that it now requires a negative energy density throughout space, asymptotically approaching zero at infinity.

An obvious alternative is to imagine that some basic momentum transfer occurs, where mass is shed in the process of creating the momentum flux in the bubble. In this way, a kind of rocket-like solution could be possible that cancels out the acceleration effects for passengers inside. However, this approach also presents its own problems since the bubble likely requires large amounts of matter to cancel out acceleration inside, thus requiring an even larger ejection of mass to accelerate itself which becomes quickly untenable. In addition, creating the metric itself which describes this situation, has not been done in detail before beyond simple photon rockets \cite{1983mgm..conf.1001W}.

Insight might be gathered by considering the ADM momentum \cite{2008GReGr..40.1997A} where conservation of the 4-momentum might be a key element to understanding the constraint to warp solutions when creating physically accelerating solutions with ADM mass. The key question in this regard is whether the `spinning-up' of the warp drive results in the forward motion of the entire structure without the need for any energy ejection. Analyzing non-vacuum spacetimes with non-Schwarzschild boundary conditions might yield valuable insight.

Another alternative is to explore the use of focused gravitational radiation emission as a way to accelerate drives over traditional momentum transfer methods, such as recently discussed in \cite{2022PhRvL.128s1102V}. In the work here, we assumed that $du_i/dt = 0$ in the passenger region during the acceleration phase of the warp drive, but this is not a requirement in general solutions. It is possible that the presence of non-zero shift vector may not be the key source of geodesic transport in all solutions if carefully constructed spatially varying lapse and metric spatial terms exist in the passenger volume. In fact, in the scenario of ejecting matter, the lapse and spatial terms will vary in time and spatially across a given spatial slice. Ultimately, the question of how to make physical and efficient acceleration is one of the foremost problems in warp drive research.

\section{Conclusion}\label{sec:conclusion}
In this paper, we have developed the first constant velocity subluminal physical warp drive solution to date that is fully consistent with the geodesic transport properties of the Alcubierre metric. It provides geodesic transport of observers while also satisfying the NEC, WEC, DEC, and SEC. This solution was constructed from a stable shell of matter with a modified shift vector on its interior, creating a warp solution with positive ADM mass. Analysis and construction of the shell used a new numerical toolkit called Warp Factory, which was developed by APL for warp research. This exciting new result offers an important first step toward understanding what makes physical warp solutions. Moreover, the warp drive spacetime constructed here is a new type of warp drive beyond the Natario class and hence not subject to the same scope discussed in \cite{2021CQGra..38o5020F} and \cite{2023arXiv230910072S} due to its use of modified spatial terms in the metric. This new solution shows that a more generic constant velocity warp drive spacetime can be constructed that satisfies the energy conditions.

We intend to explore this solution further and find areas of optimization to improve the mass-to-velocity ratio required to maintain physicality. The metric construction process of smoothing can be replaced by a direct 1D optimization of the radial profiles for density, pressure, and shift vector, possibly reducing required mass by orders of magnitude. In addition, the question of accelerating the drive efficiently without breaking physicality is a major direction of work for the field of warp drive research. The code for the metrics and figures shown here will be provided as an update to the Warp Factory codebase. 

\medskip

\noindent Warp Factory can be found at \url{https://github.com/NerdsWithAttitudes/WarpFactory}.

\clearpage

\appendix

\clearpage

\section{Numerical Error Analysis}\label{apx:erroranalysis}
Using numerical methods for analysis puts constraints on the accuracy of the results due to limitations in finite differencing methods for solving the field equations, representing the spacetime with precision-limited numbers, and discretizing the grid. These errors are summarized below: 
 \begin{enumerate}
     \item \textbf{Spherical to Cartesian Interpolation Error:}
The conversion of the spherical metric to the Cartesian metric uses Legendre polynomials to interpolate points in $r$ to points in $x$, $y$, and $z$. This interpolation introduces errors in the final metric.

\item \textbf{Finite Difference Discretization Error:}
This error comes about from the discretization of the space into a grid. With lower spatial resolution, the finite difference methods deviate from the true analytical derivatives since the step size of the finite difference algorithm is larger. This error is largest when $f(x+h)-f(x)$ is large compared to the step size $h$.

\item \textbf{Floating Point Round-Off Error:}
The numerical calculations are done in double precision. This restricts the maximum possible range of floating point values to about 16 orders of magnitude. The solver for the Einstein Field Equations is written to reduce catastrophic cancellation of small numbers, but the double precision limit still restricts meaningful results for the stress-energy tensor and energy condition violation to those of above $10^{34}$ in magnitude.

\item \textbf{Finite Difference Truncation Error:}
Finally, finite difference truncation error happens when the infinite series that calculates the derivatives is cut off to the fourth order. For the fourth-order finite difference method, the truncation error is below the double precision round-off error floor.
 \end{enumerate}
 
\begin{figure}[hbt!]
\centering
\includegraphics[width=\textwidth]{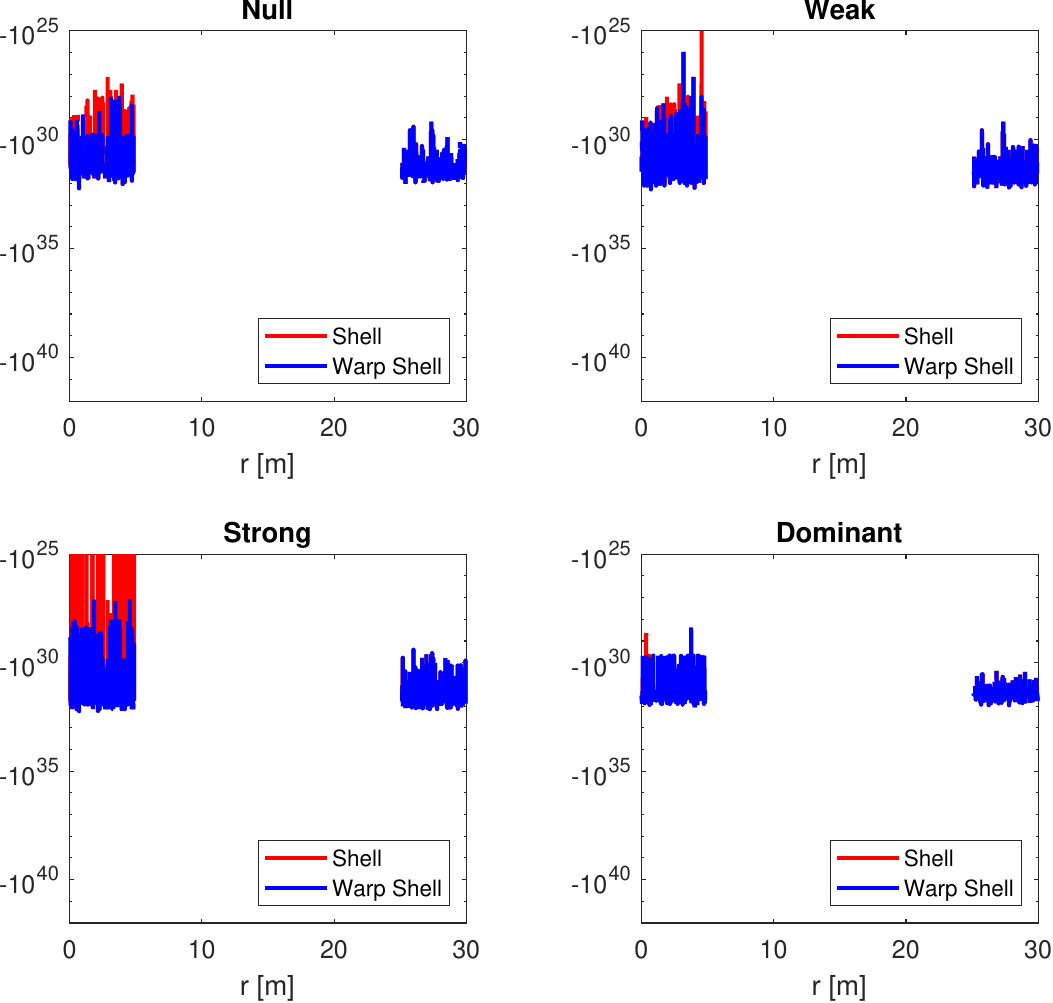}
\caption{Energy condition evaluation all the way to below the double precision floor for the Shell and Warp Shell. Only the violating values are shown. The region between $R_1$ and $R_2$ is empty as no violation exists and only positive values for all observers are found. No systematic deviation in errors is seen between the metrics. The values of the stress-energy tensor in this work are on the order of $10^{39}$, which leaves a difference between the noise floor of 
 around $10^{-6}$.}\label{fig:energyConditionError}
\end{figure}

As an example of the error floor in this analysis, we can look at the full returned values of the energy conditions in Figure \ref{fig:energyConditionError} for the Shell, Boosted Shell, and Warp Shell.

\clearpage

\section{Lorentz Transformation}\label{apx:lorentz}
The Lorentz factor is given in the usual manner:
\begin{equation}
    \gamma = \frac{1}{\sqrt{(1-\beta^2)}}
\end{equation}
Applying the Lorentz transformation corresponding to a boost along the positive x-dimension to a comoving metric $g$ results in the new metric $g^\prime$ in terms of the old components as:
\begin{align}
    g'_{00} &=  \gamma^2 \left(g_{00} - 2\beta g_{01} + \beta^2 g_{11}\right) \\
    g'_{01} &= \gamma^2 \left( g_{01} - \beta g_{11} - \beta g_{00} + \beta^2 g_{01} \right) \\
    g'_{10} &= g'_{01} \\
    g'_{02} &= \gamma \left( g_{02} - \beta g_{12} \right) \\
    g'_{20} &= g'_{02} \\
    g'_{03} &= \gamma \left( g_{03} - \beta g_{13} \right) \\
    g'_{30} &= g'_{03} \\
    g'_{11} &= \gamma^2 \left( g_{11} +  \beta^2 g_{00} - 2 \beta g_{01} \right) \\
    g'_{12} &= \gamma \left( g_{12} - \beta g_{02} \right) \\
    g'_{21} &= g'_{12} \\
    g'_{13} &= \gamma \left( g_{13} - \beta g_{03} \right) \\
    g'_{31} &= g'_{13} \\
    g'_{22} &= g_{22} \\
    g'_{33} &= g_{33} 
\end{align}
The direction of the transformation is opposite to the direction of $\beta$.

\clearpage

\section*{Biblography}

\end{document}